\documentclass[12pt]{article}
\usepackage{amsmath,amssymb,amsfonts,epsfig,graphicx}

\newcommand{\bse}{\begin{subequations}}
\newcommand{\ese}{\end{subequations}}
\newcommand{\be}{\begin{equation}}
\newcommand{\ee}{\end{equation}}
\newcommand{\bea}{\begin{eqnarray}}
\newcommand{\eea}{\end{eqnarray}}
\newcommand{\ba}{\begin{array}}
\newcommand{\ea}{\end{array}}

\makeatletter \@addtoreset{equation}{section}

\makeatother

\def\A0{{\cal A}_0}

\def\N{{\cal N}}
\def\O{{\cal O}}

\def\L5{{\cal L}_5}

\def\Tr{{\rm Tr\ }}

\def\ie{{\it i.e. }}
\def\zav{\langle z_0 \rangle}
\def\half{\frac{1}{2}}
\def\xplane{$(x_1,x_2)$ plane}
\begin{document}
\baselineskip 18pt%

\begin{titlepage}
\vspace*{1mm}%
\hfill%
\vbox{
    \halign{#\hfil        \cr
           IPM/P-2006/015 \cr
           hep-th/0602270 \cr
           } 
      }  
\vspace*{15mm}%

\centerline{{\Large {\bf On Classification Of The Bubbling Geometries}}}
\begin{center}
{\bf A.E. Mosaffa, M. M. Sheikh-Jabbari}%
\vspace*{0.4cm}

{\it {Institute for Studies in Theoretical Physics and Mathematics (IPM)\\
P.O.Box 19395-5531, Tehran, IRAN}}\\
{E-mails: {\tt mosaffa, jabbari@theory.ipm.ac.ir}}%
\vspace*{1.5cm}
\end{center}

\begin{center}{\bf Abstract}\end{center}
\begin{quote}
In this paper we classify the ten dimensional half BPS solutions of the type IIB supergravity which have $SO(4)\times SO(4)\times U(1)$ isometry found by Lin-Lunin-Maldacena (LLM).
Our classification is based on their asymptotic behavior and causal structure according which
they fall into  two classes: {\it 1)} those with $R\times S^3$ boundary and {\it 2)} those with one
dimensional light-like boundary. Each class can be divided into some subclasses depending on the asymptotic characteristics of the solutions, which in part specify the global charges defining the geometry. We analyze each of these classes in some detail
 and elaborate on their dual gauge theory description. In particular, we show that the Matrix Chern-Simons theory which is the  gauge theory dual to the LLM geometries, can be obtained as the effective theory of  spherical threebrane probes in the half BPS sector.

%

\end{quote}%
\end{titlepage}
\section{Introduction}%
According to the AdS/CFT duality there is a one-to-one correspondence between the deformations of an ${\cal N}=4,\ D=4$ supersymmetric Yang-Mills (SYM) by gauge invariant operators and the deformations of the $AdS_5\times S^5$ geometry \cite{MAGOO}. On the gauge theory side the operators are specified by their $SO(4,2)\times SO(6)$ quantum numbers as well as the number of traces. In the gravity side again the deformations can be labeled by their representation under the $SO(4,2)\times SO(6)$ isometry of the AdS background. In this picture our analysis
is usually limited only to ``small deformations'' where we can treat the deformations as
perturbation. (In the gravity side this means that we are ignoring the back-reaction on the geometry.)  In order to obtain a complete picture of the gravity/gauge theory duality we need, however, to know about the back-reactions and go beyond the perturbative description.
Although generically very hard, this has been done for some specific deformations.

In \cite{LLM}, Lin-Lunin-Maldacena (LLM) constructed the gravity solutions corresponding to
all 1/2 BPS deformations of the ${\cal N}=4$ SYM on $R\times S^3$, that is deformations of the SYM by the chiral primary operators. In the half BPS sector the operators are determined by a
single quantum number, the R-charge $J$, which is equal to their scaling dimension $\Delta$.
As such the chiral primary operators are singlets of $SO(4)\times SO(4)\times U(1)\in SO(4,2)\times SO(6)$.
A chiral primary operator with R-charge $J$ is then completely specified if we determine
how the $J$ chiral fields of the ${\cal N}=4$ SYM $Z$ are distributed in various traces.
Being   $SO(4)\times SO(4)$ invariant and also noting that their scaling dimension is protected by supersymmetry one can argue that these deformations may be described by a $2d$ fermion  system \cite{Beren1, Antal}. This system may also be understood as a specific quantum Hall system with filling factor equal to one which has a manifest particle-quasihole symmetry \cite{Beren2, LLMQHS, LLLvsLLM}.

The LLM geometries preserve 16 supersymmetries which form a $PSU(2|2)\times PSU(2|2)\times U(1)$
superalgebra (for a review on these algebras see e.g. \cite{review}). This supergroup is a subgroup of $PSU(2,2|4)$ algebra, the superisometries of the $AdS_5\times S^5$ geometry.
Although the supersymmetry of the LLM geometries is a subgroup of $PSU(2,2|4)$, the LLM geometries are not generically (small) deformations of the $AdS_5\times S^5$ and they may have a different causal structure. The purpose of this paper is to classify the LLM geometries by their casual structure and asymptotic behavior.

The LLM geometries are solutions of type IIB supergravity given by \cite{LLM}
\begin{equation}\label{LLM-geometry}
ds^2=-h^{-2}\left(dt+V_i dx^i\right)^2+ h^2\left(dy^2+dx_i^2\right)+ye^{-G}d\Omega_3^2+
 ye^{G}d\tilde\Omega_3^2\ ,
\ee
with a constant dilaton and a selfdual RR fiveform turned on and
\be\label{V,G,z}
\begin{split}
h^{-2}&=2y\cosh G\ ,\ \ \ \ \ \ \  z=\frac{1}{2} \tanh G\ ,\cr
y\partial_y V_i&=\epsilon_{ij} \partial_j z\ ,\ \ \ \ \ \ \ \epsilon_{ij}\partial_iV_j=\frac{1}{y}\partial_y z\ .
\end{split}
\ee
As we see the whole solution is determined through a single function $z=z(y;x_i),\ i=1,2$. From
\eqref{V,G,z} it is inferred that
\be\label{Laplace}
\partial_i^2 z+{y} \partial_y\left(\frac{1}{y}\partial_y z\right)=0\ .
\ee
In other words $\frac{z}{y^2}$ satisfies a six dimensional Laplace equation. Demanding the smoothness of the solutions restricts the function $z$ at $y=0$ to only take values $\pm \frac{1}{2}$. The solutions to \eqref{Laplace} are then determined by the values of $z$ at $y=0$ as the source. That is \cite{LLM}
\be\label{z-sol'n}
z=\frac{y^2}{\pi}\int d^2x'\ z(0;x'_i)\frac{1}{\left(y^2+(x_i-x'_i)^2\right)^2}\ .
\ee
Therefore, as it is customary, a generic LLM solution can be specified by a black (white) color-coding attributed to
$z=-1/2\ (+1/2)$ regions on the $(x_1,x_2)$ plane.

The above smoothness condition is not complete unless we add the quantization of the area on the
$(x_1,x_2)$ plane, explicitly \cite{LLLvsLLM, Alishah}
\be\label{NCy}
[x_1,x_2]=2\pi i l^4_p\ .
\ee
That is, the  $(x_1,x_2)$ plane is a Moyal plane in which the area of both black and white regions is quantized.
This quantization leads to the quantization of the fiveform flux in the supergravity level
\cite{LLM}.

In this paper we continue the analysis of the LLM geometries,  classifying them by their causal structure and the large $y$ behavior of the $z$ function. In section 2, we show that the causal boundary of the LLM geometries are only specified by the average of $z_0=z(y=0)$ over the $(x_1,x_2)$ plane, which will be denoted by $\langle z_0 \rangle$.
One can then distinguish two distinct cases $\zav =\pm \frac{1}{2}$ and $\zav \neq\pm \frac{1}{2}$. In the former case the boundary is four dimensional $R\times S^3$ and in the latter it is one dimensional light-like. In sections 3 and 4, we refine this classification by considering various moments of the black and white distribution. In section 3, we consider the
$\zav =\pm \frac{1}{2}$ cases where one can distinguish two classes with
finite and infinite area of the black region. In both cases this is the zeroth and second moments of the distribution which is relevant. In section 4, we study the
 $\zav \neq\pm \frac{1}{2}$ cases and discuss that the classification maybe refined by the first and the zeroth moments of the distribution. In sections 3 and 4, we also discuss the dual field theories to each of these cases separately, which are all related to noncommutative Matrix Chern-Simons theory. We show how this Matrix Chern-Simons theory can be obtained from the effective action of
half BPS spherical three-branes probing the LLM geometry. The last section is devoted to discussion and outlook.

\section{Causal Structure of the LLM Geometries}
In this section we classify the bubbling geometries with their causal structure. For this, we investigate the existence of a causal boundary for the geometries and then relate the structure of the boundary to the properties of the function $z$ on $(x_1,x_2)$ plane. Our classification of the causal structure will thus become a classification of the different behaviors $z$ can have on the plane. This goes in line with the very important feature of these solutions according which the whole geometry is obtained by the value of $z$ on the \xplane . To simplify the discussion, we make use of the $Z_2$ symmetry of the LLM solutions which interchanges black and white boundary conditions \cite{Alishah} and therefore we can restrict our attention to the situation where $0\le z\le1/2$. The complementary range $-1/2\le z\le 0$ can be reached from the former by the mentioned $Z_2$ action.

The upshot of our analysis is the following statement:
\begin{center}
The average value of $z$ on the \xplane, $\zav$, determines the causal structure.\\
For $\zav=1/2$ the boundary is $R\times S^3$ and for $\zav\ne 1/2$ it is one dimensional light like.
\end{center}

To prove this statement we define the parameter $\theta$ by $\tan\theta=e^{-G}$ in terms of which
\be\label{z-theta}
z=\frac{1}{2}\tanh G=\frac{1}{2} \cos 2\theta\ ,
\ee
and restrict ourselves to $0\leq z\leq \frac{1}{2}$, $\theta\in [0,\frac{\pi}{2}]$.
Using $\theta$ instead of $G$, the LLM ansatz finds a more illuminating form for the current discussion. Moreover, we use polar coordinates $(r,\alpha,\beta)$ for the space $(x_1,x_2,y)$ with the usual definition
\be
y=r\cos\alpha\,,\,\,\,\,\,x_1=r\sin\alpha\cos\beta\,,\,\,\,\,\,x_2=r\sin\alpha\sin\beta\,.
\ee
In terms of these variables the LLM metric is written as
\be
ds^2=\frac{2r\cos\alpha}{\sin2\theta}[-(dt+V_rdr+V_\gamma d\gamma)^2+\frac{\sin^22\theta}{4\cos^2\alpha}(\frac{dr^2}{r^2}+d\alpha^2+\sin^2\alpha d\beta^2)+\sin^2\theta d\Omega_3^2+\cos^2\theta d\tilde{\Omega}_3^2]\ ,
\ee
where $\gamma=\alpha,\beta$.\footnote{Since in the $x_1,x_2,y$ coordinate system $V$ only has $V_{x_1}, V_{x_2}$ components, $V_r$ and $V_\alpha$ components are not independent and related as $V_\alpha=r\cot\alpha V_r$. $V_r, V_\beta$ in terms of $V_1, V_2$ are then given by $V_r=\sin\alpha(V_1\cos\beta +V_2\sin\beta )$ and
$V_\beta=r\sin\alpha(V_2\cos\beta-V_1\sin\beta)$.}
 Note that ({\it cf.} \eqref{z-theta}) $\sin2\theta$ can only vanish when $z=1/2$. The causal structure (Penrose diagram) of this geometry can be determined if one can bring the metric into a form which is conformally an Einstein static Universe with all the spatial coordinates having a finite range.

In the case of generic 10 dimensional LLM geometries, the Penrose diagram is generically a six dimensional diagram and one cannot suppress more dimensions therefore it will not be instructive to show the Penrose diagram. Hence, we will only focus on extracting the
 structure of the causal boundary which, recalling that LLM geometries are {\it non-singular, smooth and have no horizons}, is the only interesting information contained in the Penrose diagram.

 The causal boundary is the locus which is not formally a part of our space-time, but in causal contact with all the points in the geometry. That is, it is a place where one can send
and receive light rays in finite coordinate time. In the coordinates where metric is conformal to Einstein static Universe the points where the conformal factor blows up determine the locus of the causal boundary.
Let us first see whether the above form for the metric serves this requirement or we have to pull another conformal factor out of the expression in the brackets.

The conformal factor $\frac{2r\cos\alpha}{\sin2\theta}$ can blow up either if $\sin2\theta=0$ or $r\cos\alpha$ goes to infinity. The former, in turn, can happen either on the  \xplane\ or somewhere at $y\ne0$. But it should be noted that as one approaches $y=0\ (\alpha=\pi/2)$, $z$ behaves as $z\sim 1/2-f(x)y^2$. From this it follows that, at finite $x_i$, $\sin2\theta\sim y$ and therefore the conformal factor behaves as $\frac{2r\cos\alpha}{\sin2\theta}\sim 1$ and can never blow up on the plane. Outside the \xplane, however, $\sin2\theta$ can approach zero only if $\zav=1/2$ and the limit is reached as $y$ goes to infinity. It therefore follows that for configurations with $\zav=1/2$ the  above two possibilities $(r\cos\alpha\rightarrow\infty \,\rm{and}\, \sin2\theta=0)$ coincide. So for $\zav=\frac{1}{2}$ configurations and as long as the causal structure is concerned what matters is the large $r$ behavior  where
\be\label{large-r-z1/2}
z\sim\frac{1}{2}-\frac{1}{r^2}\,,\,\,\,\,\,\sin2\theta\sim\frac{1}{r}\,,\,\,\,\,\,V_r\sim\frac{1}{r^2}\,,\,\,\,\,\,V_\gamma\sim\frac{1}{r}\ .
\ee
One can now write the asymptotic form of the metric
\[
ds^2=\frac{2\cos\alpha}{\rho\sin2\theta}[-dt^2+Ad\rho^2+Bd\rho dt+\frac{\sin^22\theta}{4\cos^2\alpha}(d\alpha^2+\sin^2\alpha d\beta^2)+\sin^2\theta d\Omega_3^2+\cos^2\theta d\tilde{\Omega}_3^2+{\cal O}(\rho)]\ ,
\]
where 
\[
\rho=\frac{1}{r}\ ,
\]
and $A, B$ are only functions of $\alpha$ and $\beta$ with no $\rho$ dependence. Note that as $r$ goes from a minimum value $r_{min}$ to infinity, $\rho$ covers a finite range and the above form of the metric has the desired properties for studying the causal structure. Now at $\rho=0$, the conformal factor blows up and since $\sin2\theta=0$ at this point, either $\sin\theta$ or $\cos\theta$  becomes zero. Therefore, in either case the radius of one of the
three spheres vanishes and what remains in the bracket is
\be\label{S3,tildeS3}
-dt^2+d\Omega_3^2\,\,\,\,\,\,\,\rm{or}\,\,\,\,\,\,\,-dt^2+d\tilde{\Omega}_3^2\ ,
\ee
which describes the boundary of the space time, because one can send a light ray along the $\rho$ direction from a finite $\rho_0$ to $\rho=0
\ (r=\infty)$ in a finite coordinate time $t$.
The final result is that those LLM geometries which are specified by black and white configurations on $(x_1,x_2)$ plane with $\zav=1/2$, have a causal boundary of the form $R\times S^3$.

Now let us consider the second possibility for the conformal factor to blow up \ie $r\cos\alpha\rightarrow\infty$ with $\sin2\theta\ne0$. This can happen for configurations with $0\le \zav <1/2$ for which far from the $(x_1,x_2)$ plane
\be\label{large-r-zn1/2}
z\sim \zav-\frac{1}{r^n}\,,\,\,\,\,\,\sin2\theta\sim 1-4\zav^2+\frac{1}{r^n}\,,\,\,\,\,\,V_r\sim\frac{1}{r}\,,\,\,\,\,\,V_\gamma\sim1\ ,
\ee
where $n$ is a positive number, in the section 4 we will discuss several examples with $n=1, 2$ and as we will see momentarily the causal structure is independent of the value of $n$.

The asymptotic form of the metric in the large $r$ can be  written as
\[
ds^2=\frac{2e^\rho\cos\alpha}{\sin2\theta}[-dt^2+Ad\rho^2+Bd\rho dt+\frac{\sin^22\theta}{4\cos^2\alpha}(d\alpha^2+\sin^2\alpha d\beta^2)+\sin^2\theta d\Omega_3^2+\cos^2\theta d\tilde{\Omega}_3^2+{\cal O}(e^{-\rho})]\ ,
\]
where
\[
\rho=\ln r\ ,
\]
and
\[
B=2rV_r\ ,\ \ \ \ \ A=-\frac{B^2}{4}+\frac{\sin^2 2\theta}{4\cos^2\alpha}\ ,
\]
are functions of $\alpha$ and $\beta$ with no $\rho$ dependence. The problem now is that $\rho$  still has an infinite range  as $r$ goes from $r_{min}$ to infinity and hence we have to pull out another conformal factor from the bracket. This can be done by the following change of variables
\be
t+{C_\pm}{\rho}=\tan(\frac{\psi\pm\xi}{2})\,,\,\,\,\,\,C_\pm={\frac{B}{2}\pm\frac{\sin2\theta}{2\cos\alpha}}\ ,
\ee
in terms of which the metric is written as
\[
ds^2\sim\frac{2\exp({\frac{\sin\xi}{\cos\psi+\cos\xi}})\cos\alpha}{\sin2\theta}\frac{1}{4\cos^2(\frac{\psi+\xi}{2})\cos^2(\frac{\psi-\xi}{2})}[-d\psi^2+d\xi^2+4\cos^2(\frac{\psi+\xi}{2})\cos^2(\frac{\psi-\xi}{2})(\cdots)]\ ,
\]
where $\cdots$ shows the two three spheres.
We can now safely discuss the causal structure of the above metric. The conformal factor blows up if $\psi\pm\xi=\pi$. For  either choices of the sign the radius of both of the three spheres vanishes and  what remains in the bracket is just
\[
-d\psi^2+d\xi^2\ ,
\]
which restricting to $\psi\pm\xi=\pi$, describes a null curve.  The causal boundary  is thus a one dimensional light-like space. As the final result, the LLM geometries which are specified by  configurations on $(x_1,x_2)$ plane with $0\le \zav <1/2$ have a one dimensional light-like boundary.

In light of the above analysis, there are several comments in order:

$\bullet$ Only the average value of the function $z$ on the $(x_1,x_2)$ plane, $\zav$, classifies the bubbling geometries in terms of their causal structure and with respect to this property the geometries fall into two classes. Out of the whole range that the average can take, $0\le \zav \le1/2$, the point $\zav=1/2$ is singled out which constitutes one of the two classes \ie geometries with $R\times S^3$ as the causal boundary. The complementary range, $0\le \zav <1/2$, constitutes the other class \ie those with a one dimensional light-like boundary. In the former case always one of the three spheres shrinks to a point and the remaining one constitutes the compact part of the boundary whereas in the latter case both three spheres shrink.

$\bullet$ The difference between the two $\zav=1/2$, $\zav \neq 1/2$ cases stems from   the large $r$ asymptotic behavior of $\sin2 \theta$. For the former $r\sin 2\theta\sim 1$ ({\it cf.} \eqref{large-r-z1/2}) while in the latter
$\sin2 \theta\sim 1-4\zav^2\neq 0$ ({\it cf.} \eqref{large-r-zn1/2}).

$\bullet$ In either $\zav=1/2$ and $\zav\neq 1/2$ cases, the boundary is never along the subspace $(x_1,x_2,y)$.

$\bullet$ As two famous examples of the two cases, one can mention the $AdS_5\times S^5$ geometry
which has $\zav=1/2$ and the ten dimensional maximally supersymmetric plane-wave
which has $\zav=0$. The former has a four dimensional boundary $R\times S^3$ \cite{MAGOO}
and the latter a one dimensional null boundary \cite{review, BN}.

$\bullet$ Although the two $\zav=1/2$ and $\zav=-1/2$ both have $R\times S^3$ as boundary,
the three spheres which appear along the boundary are different, for the
former it is $R\times S^3$ and for the latter $R\times {\tilde S}^3$ ({\it cf.} \eqref{S3,tildeS3}). This could
be understood easily noting the $Z_2$ symmetry discussed in \cite{Alishah}.

$\bullet$ As we discussed the causal structure only depends on the large $r$ (large $y$ or large $x_i$)
behavior of the $z$-function and on the other hand it is given by the average value of $z$
at $y=0$. It is desirable to have everything in a uniform language. This is possible noting
the fact that average value of $z$ on the \xplane\ at $y=0$ is equal to the average value of $z$
at $y=\infty$, \ie
\be
\langle z\rangle_{y=0}= \langle z\rangle_{y=\infty}\ .
\ee
This can directly be confirmed using the equation \eqref{z-sol'n}.
In particular we note that  $z$ cannot take values $\pm\half$ anywhere at finite $y\neq 0$ \cite{Caldarelli, Olaughlin} and since it asymptotes to $\langle z\rangle$, it can only acquire $\pm1/2$ values at non-zero $y$, if $\langle z\rangle=\pm\frac{1}{2}$ and this can only happen at $y=\infty$.

$\bullet$ The above results are reasonable and expected when we consider continuous deformations of the black and white distribution on the $(x_1,x_2)$ plane. Firstly, we expect that the causal structure remains unchanged under such {\it finite} smooth continuous deformations. Secondly, as can be easily conceived, the point $\zav=1/2$ is again singled out as it is a fixed point for the finite deformations. The range $0\le \zav <1/2$, however, can be covered continuously by the deformations and hence we expect that the corresponding geometries have identical causal structures. So once the connection between $\zav$ and the causal structure is established, from this simple argument one can both identify the two classes of geometries and also identify the causal boundaries in each class by looking at well understood examples in either case, say, (asymptotically) $AdS_5\times S^5$ for the first class and plane-waves for the second.

\section{LLM geometries with $R\times S^3$ as the boundary}

In this section we discuss the first class of geometries mentioned in the previous section \ie those which have $R\times S^3$ as the causal boundary. As mentioned before, such geometries must be described by a $z$ function with $\zav=1/2$. These geometries fall into two classes themselves. The first class constitutes of geometries which are finite deformations of, and asymptote to $AdS_5\times S^5$. A generic example of such solutions is a collection of concentric rings around a circular droplet as the $z$ configuration on the $y=0$ plane. In this class of solutions the black areas on the boundary plane are confined in a finite region and have a limited extent and thus the $\zav=1/2$ requirement is trivially satisfied. As the second class one can consider black areas on the  \xplane\ having an infinite extent in such a way that the ratio of total black to white area  is zero. A generic example of such configurations is a collection of black strips. Examples of these two cases have been depicted in Fig.\ref{Rings-Fig}.
In the following we discuss these two cases separately by focusing on the rings and strips examples.
\begin{figure}[ht]
\begin{center}
\includegraphics[scale=0.7]{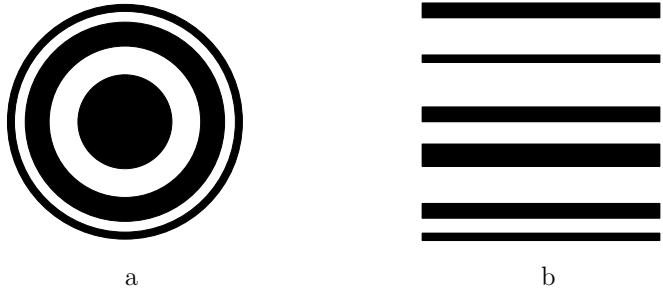}\caption{Two generic examples for cases with $\zav=1/2$. (a) A collection of concentric rings with finite extent and (b) a collection of strips with an infinite extent of black areas.}
\label{Rings-Fig}
\end{center}
\end{figure}

\subsection{The case with finite area of the black region}

Let us first study geometries which are asymptotically $AdS_5\times S^5$, an example  of which has been depicted in Fig. \ref{Rings-Fig}(a). On the gravity side, these geometries are characterized by  quantum numbers which can be identified with the moments of the distribution on the  \xplane, $z_0=z(x_1,x_2; y=0)$. The zeroth moment,
\be\label{N-def}
N\equiv\frac{-1}{4\pi^2 l^4_p}\int d^2x\ (z_0-\frac{1}{2})=\frac{1}{4\pi^2 l^4_p}\int_{{\cal D}} d^2x  \ ,
\ee
is the total black area and is finite in this case. Note that the first integral is over the whole \xplane\ while the second is only on the black region, the ``droplet''. For the finite $N$ case the first moment can always be set to zero  by an appropriate choice of the origin of the coordinate system and the second moment $J$,
\be\label{J-def}
J\equiv\frac{1}{16 \pi^3 l_p^8}\left[\int_{{\cal D}} d^2x\ (x_1^2+x_2^2)-\frac{1}{2\pi}\biggl(
\int_{{\cal D}} d^2x\biggr)^2\right]\ ,
\ee
is related to  the angular momentum of the solution \cite{LLM}. The higher moments describe the details of the distribution and are not related to global charges of the geometry.

These data, and in particular $N,J$,  can also be read from  the large $y$ behavior of $z$ and this is what we do in the following. Consider the expression for $z$ \eqref{z-sol'n} which can be written as
\be\label{z-droplet-in}
z(x_1,x_2,y)=\frac{-y^2}{\pi}\int_{\cal{D}}\frac{dx'_1dx'_2}{[(x_i-x_i')^2+y^2]^2}+\frac{1}{2}\ ,
\ee
where the integral  is over the droplet (the black region). We now make an expansion in $1/y^2$
\be\label{expansion-case1}
\begin{split}
z(x_1,x_2,y) &=\frac{1}{2}+\frac{-1}{\pi
y^2}\big(\int_{\cal{D}} d^2x'-\frac{2}{y^2}\int_{\cal{D}}d^2x'(x_i-x_i')^2+\cdots\big)\cr
&= \frac{1}{2}+ \frac{2\pi l_p^4}{y^2}\ N+\frac{32\pi^2l_p^8}{y^4}\left[
(J+\frac{1}{8}N^2)-\frac{N}{8\pi^2l_p^4}x_i^2\right]+{\cal O}(\frac{1}{y^6})\ .
\end{split}
\ee
In this expression the leading term, $1/2$, is $\langle z\rangle_{y=\infty}$ which is, as mentioned before, equal to $\langle z\rangle_{y=0}$ and determines the causal boundary. The next order term, which is of order $\frac{1}{y^2}$, corresponds to the total area of the black regions on the plane, \ie $N$ which is one of the global charges and also a good quantum number for the configuration. The $\frac{1}{y^4}$ term determines the next quantum number $J$.
As mentioned earlier, the geometries on this case are asymptotically $AdS_5\times S^5$
geometries with $R_{AdS}^4=4\pi l_p^4 N$, which is deformed by a collection of giant gravitons
carrying angular momentum $J$.

One may ask about the dual gauge theory picture for the LLM geometries.  The sector in the ${\cal N}=4$ $U(N)$ SYM dual gauge theory which is equivalent to the above geometries is equivalently described by a system of $N$ one dimensional fermions \cite{LLM, Beren1, Antal, Nemani, Mandal} and the phase space of these
 fermions may directly be identified with the \xplane\ in the LLM geometries \cite{LLLvsLLM}.
In fact, it has been argued that the system of these fermions is equivalent to a quantum Hall system (QHS) with filling factor equal to one, a system with explicit particle-quasihole symmetry \cite{Beren2, LLMQHS, LLLvsLLM}.

As has been reviewed and discussed in some detail in \cite{LLLvsLLM}, the two dimensional
QHS can be described by a Matrix Chern-Simons theory, or a Chern-Simons theory on the
noncommutative Moyal plane. Here we show a different route to obtain the Matrix Chern-Simons theory, other than starting from the ${\cal N}=4$ $U(N)$ SYM and restricting to the sector involving only chiral primary operators. The idea is to use spherical three brane probes to probe the completely white \xplane. The number of branes we choose
is of course $N$, the area of the black region (in units of $4\pi^2 l_p^4$).
Our intuition is that the LLM geometry with black region of area $N$
is the near horizon geometry of the supergravity solution corresponding to $N$
spherical branes on the totally white \xplane\ LLM solution. Or equivalently,
back reaction of $N$ spherical three-branes on the totally white \xplane\ is
described by the LLM geometries described by droplets of area $N$.
In particular the $AdS_5\times S^5$ geometry \textit{in the global coordinates}, which is described by a black disk in the background white \xplane, is nothing but the near horizon geometry of the $N$ spherical three branes in the totally white \xplane\ background.

Our strategy is then to apply the BFSS matrix theory ideas \cite{BFSS}: the fact that M/string theory
on a background is described by  the low energy effective theory of D-branes probing that geometry while we can generically ignore the back reaction of the branes on the geometry.
Note that this is not exactly what we do in the AdS/CFT type dualities.
Here we start with the LLM geometry corresponding to the totally white \xplane\ as the background and probe it with spherical three-branes. These spherical branes are the appropriate objects for the sector we are interested in, the half BPS objects and the LLM geometries.
\footnote{Although similar ideas and using the three brane probes have been considered previously
\cite{Mandal}, our approach is different in the sense that we directly apply the BFSS matrix theory ideas.}
In fact it has been conjectured that \cite{TGMT} the spherical three-branes with unit angular momentum, the         ``tiny gravitons'', are capable of describing, not only the theory in the half
BPS sector \cite{Torab}, but also the full type IIB string theory on the plane-wave or the
$AdS_5\times S^5$ in the DLCQ description.

Consider the  totally white boundary condition on the $y=0$ plane where $z=1/2$ everywhere on the plane. This boundary condition results in a $z$ that is constant and equal to 1/2 everywhere in its $(x_1,x_2,y)$ domain  which implies that $G$ is also a constant and very large. One also obtains that $V=0$. The background reads as
\bea
ds^2&=&h^{-2}(-dt^2+d\Omega_3^2+h^4dx^idx^i)+h^2(dy^2+y^2d\tilde{\Omega}_3^2)\ ,\\
h^{-2}&=&y{\rm e}^G\ .\cr
F_{(5)}&=&\frac{1}{4}\bigg(-d(y^2{\rm e}^{2G})\wedge dt+\epsilon_{ij}dx^i\wedge dx^j\bigg)\wedge d\Omega_3\ .
\label{wb}
\eea
We choose to probe the above geometry with $N$  spherical three-branes  wrapping around $\Omega_3$ and since $G\rightarrow\infty$ we take $y\rightarrow 0$ such that $y{\rm e}^G\equiv h^{-2}$ is constant. That is, we freeze the fluctuations of the brane, as we are only interested in the half BPS configurations. Therefore, the second part of the metric becomes irrelevant in this analysis and the first term in the expression for $F_{(5)}$ vanishes. As a result, the part of the RR four form which couples to the branes is
\be
C_{(4)}=\frac{1}{4}\ \epsilon_{ij} x^idx^j\wedge d\Omega_3\ .
\ee
Assuming that the gauge and fermionic fields on the branes are not excited (which is dictated by the half BPS condition), the world volume action is written as
\be
S=-\frac{1}{g_s}Vol(\Omega_3)\int dt \  h^{-4}\ \Tr\sqrt{1-{h^4}(\dot{X}^i\dot{X}_i)}+Vol(\Omega_3)\int dt\ \frac{1}{4}\ \epsilon_{ij}\ \Tr(X^i\dot{X}^j)\ ,
\ee
where $X^i$ are $N\times N$ unitary matrices representing the collective coordinates of the $N$ probe branes. To write the above action for a collection of branes we have used
the prescriptions  of \cite{Raamsdonk}.

Next we expand  the square root, drop the overall factor of $Vol(\Omega_3)$ and absorb $g_s$ and some numeric factors in a scaling of $t$. The nontrivial part of the action becomes
\be
S=\int dt\ \Tr(\frac{1}{2}{D_0 X}^i{D_0 X}_i+\frac{1}{2} \epsilon_{ij} X^i{D_0 X}^j)\ .
\ee
In the above action, along with the arguments of \cite{Raamsdonk}, we have re-introduced the only component of the $0+1$ gauge field
$A_0$ through the covariant derivative
\[
D_0 X^i=\partial_0 X^i+i[A_0, X^i]\ .
\]
This action is nothing but the matrix version of the Landau problem \ie the problem of $N$ electric charges on a plane in a constant magnetic field with the potential $A_i=\frac{1}{2}\epsilon_{ij}x^j$. In the limit where the branes are sufficiently separated such that the matrices become diagonal, the above action exactly reduces to that for the Landau problem.

If we require the spherical branes to be BPS, we have to impose a further restriction on the above action. It is well known that this requirement amounts to reducing the action to the Chern-Simons term \ie dropping the kinetic term. In terms of the Landau problem, this is equivalent to going to the Lowest Landau Level (LLL) which is described by a Quantum Hall System (QHS) \cite{LLLvsLLM}. In the end, the dynamics of BPS spherical three branes in the background (\ref{wb}) is given by
\be
S=\int dt\ \epsilon_{ij}\Tr(\ X^iD_0{X}^j)\ .
\label{cs}
\ee

An important conceptual consequence of the above analysis is the identification of the coordinates $x^i$ with the collective coordinates of probe branes and since these are expressed in terms of matrices $X^i$, noncommutativity of the plane follows immediately. This direct link is not visible in the usual AdS/CFT guided study of the LLM geometries. Furthermore, the commutator $[X_1,X_2]$ is proportional to the inverse of the density operator for the particles
\cite{LLLvsLLM} and the Wigner function corresponding to this operator is identified with the distribution $\tilde{z}=z-\frac{1}{2}$ on the \xplane\ \cite{Mandal, Vijay, Silva}.

\subsection{The case with infinite area of the black region}

The second class of $\zav=\frac{1}{2}$ configurations that we consider are those with  infinite area of black region, \ie infinite $N$. The simplest case of this case, on which we will concentrate in this section, are those depicted Fig. \ref{Rings-Fig}(b). As the ``droplets'' have infinite extent in one direction, the quantum numbers $N$ and $J$ which are defined through  \eqref{N-def} and \eqref{J-def} characterize states in the previous case are not  relevant for these geometries. In order to read the good quantum numbers, similarly to the previous case which was done in \cite{LLM}, we analyze the large $y$ behavior of the $z$-function. The idea is to identify the ADM mass, (angular) momentum and other physical quantities of the metric. The starting point is \eqref{z-sol'n} which  noting the translation symmetry along $x_1$ leads to
\be\label{z-strips}
z=\half-\frac{y^2}{2}\int_{\cal S} dx'_2 \frac{1}{\left((x_2-x'_2)^2+y^2\right)^{3/2}}\ ,
\ee
where the integral ${\cal S}$ is over the black strips. If the distribution of the black strips along $x_2$ direction has a finite extent, one can perform a large $y$ expansion:
\be\label{z-strips-large-y}
z=\half-\frac{\Delta}{2 y}+\frac{3}{4y^3}\left[\Delta x_2^2+\frac{\Delta^3}{12}+K\right]+
\O (\frac{1}{y^{5}})\ ,
\ee
where
\begin{subequations}
\begin{align}
\Delta &\equiv \int_{\cal S} dx'_2\ ,\\
K&\equiv  \int_{\cal S} dx'_2\ {x'}_2^2-\frac{\Delta^3}{12}\ ,
\end{align}
\end{subequations}
and we have chosen the origin so that the first moment is zero, \ie $\int_{\cal S} dx'_2\ x'_2=0$. $\Delta$ and $K$, which are
respectively the zeroth and the second moment of the distribution of the strips, are among the quantum numbers which  describe the solutions of this case.

From \eqref{z-strips-large-y} it is evident that the average of $z$ is $\half$, however, the solution is not asymptotically AdS. This can be seen from the next leading term which unlike the asymptotically AdS case, goes as $\frac{1}{y}$. As we'll see momentarily the quantum numbers
$\Delta, K$ respectively play the role of $N, J$ in the AdS case.

In this case the metric  has a translational symmetry along the $x_1$ direction and one can compactify the $x_1$ direction on a circle of radius $R$. The \xplane\ then becomes a noncommutative cylinder  and hence the spectrum of the $x_2$ becomes discrete (e.g. see \cite{Chaichian}), \ie the width
of black  (or white) strips is an integer multiple of $\frac{2\pi^2 l_p^4}{R}$:
 \be\label{width-quantized}
\Delta= \frac{2\pi^2 l_p^4}{R} \ k,\ \ \ \ k\in \mathbb{Z}\ .
\ee
(Note that in our units, $x_1,x_2, y$ and hence $\Delta, R$ all are of dimension of length squared.)

Let us consider the single strip case of width $\Delta$. For this case
the quantum number $K$ vanishes. In the large $y$ limit
\footnote{To obtain the metric we also need the $V_i$'s which are given by
\[
V_2=0\ , \ \ V_1 =-\frac{1}{2}\int_{\cal S} dx'_2 \frac{x_2-x'_2}{\left((x_2-x'_2)^2+y^2\right)^{3/2}}
= -\frac{1}{2}\frac{x_2}{y^3}\Delta+\O(\frac{1}{y^5})\ .
\]}
\be\label{large-y-single-strip}
\begin{split}
ds^2&=ye^G\left[-dt^2+ d\Omega^2_3\right]+(ye^G)^{-1}\left[dx_1^2+dx_2^2+dy^2+y^2d{\tilde\Omega}_3^2\right]\\
&= f^{-1}\left[-dt^2+ d\Omega^2_3\right]+f\left[dx_1^2+dr^2+r^2d{\tilde\Omega}_4^2\right],
\end{split}
\ee
where $r^2=x_2^2+y^2$ and $f=f(r)=(ye^{G})^2=\frac{\Delta}{r^{3}}$.  The above metric is
the solution corresponding to the near horizon limit of $k\propto \Delta$ ({\it cf.} \eqref{width-quantized}) coincident spherical three branes (giant gravitons) which are uniformly smeared along the $x_1$ direction. As we see in this limit, and for the single strip case,  the $SO(4)\times SO(4)$
isometry is enhanced to $SO(4)\times SO(5)$. One may now make a T-duality along the $x_1$ direction, where the solution becomes that of $k$ coincident D4-branes with the worldvolume
along $t,x_1, \Omega_3$. The low energy effective theory is then a supersymmetric $U(k)$ $4+1$ dimensional gauge theory on $R^{1,1}\times S^3$ \cite{LM}. The action of this gauge theory besides
the Yang-Mills part also contains a term coming from the Chern-Simons piece showing the coupling of the brane to the background RR four-form field strength; this additional term can be worked out using results of \cite{Raamsdonk}, as we did in the previous subsection. In the half BPS sector one should then turn off the gauge fields along the $S^3$ and the scalar fluctuations along $\tilde S^3$. This leads to the effective $1+1$ dimensional $U(k)$ gauge theory:
\[
S=\int d^2x Tr\left[F_{\mu\nu}^2+ (D_\mu X_2)^2+ \epsilon^{\mu\nu} F_{\mu\nu} X_2\right]\ .
\]
where $X_2$ is the scalar field corresponding to the fluctuations of the branes along the $x_2$ direction.  To restrict the above action to the half BPS sector one still needs to impose a condition, which parallels that of going to the lowest Landau level in the related quantum Hall problem discussed in the previous subsection. That is, in the half BPS sector one can drop the first two terms and remain with the last.

One may consider a collection of strips of  width $\Delta_i$. The T-dual of the asymptotic
form of the metric is that of stacks of $k_i$ number of D4-branes separated along the $x_2$ direction, and hence the dual field theory is a generalization of the above gauge theory
to $\prod_i U(k_i)$ theory, which can in part be understood as a $U(\sum_i k_i)$ gauge theory Higgsed down to   $\prod_i U(k_i)$. The quantum number $K$ is then related to the overall
characteristic of the Higgsing. This theory may also be uplifted to M-theory as the half BPS
sector of the mass deformed $D=3, \N=8$ SCFT theory and the corresponding Bena-Warner supergravity solution \cite{BW}. As these theories have been studied in some detail in
\cite{LLM, LM} here we do not analyze them further.

\section{LLM geometries with one dimensional null boundary}

In this section we elaborate more on the LLM geometries with $\zav\neq \frac{1}{2}$.
One may recognize several different sub-classes, all of which are common in the fact that, in order to have $\zav \neq \frac{1}{2}$,  the black region should be extended off to infinity.
That is, they all come with $N\to \infty$. For the same reason the second moment $J$ also goes to infinity. Therefore, for these solutions one should find other good quantum numbers.

Here we only focus on  three interesting cases which have different qualitative behavior and introduce good quantum numbers for each case. Our guiding criterion for distinguishing these classes is the symmetry of the distribution $z$ on the \xplane. Given a two dimensional plane we can have translational, rotational and scaling symmetries which could be used as a basis for distinguishing various cases.
The three cases which we consider are those which are asymptotically plane-wave, these have translational symmetry along $x_1$ direction; those which have ``scaling symmetry'' in the \xplane\ and finally those
in which  \xplane\ is wrapping a two tours.  The first case has been discussed in some detail in \cite{LLM} and the follow-up papers and we will be very brief on that. The latter two cases have been previously considered
in \cite{Shepard, LM} and here we will analyze some other aspects of them.
An example of each of these cases has been depicted in Fig. \ref{Four-Figs}.
\begin{figure}[ht]
\begin{center}
\includegraphics[scale=1.3]{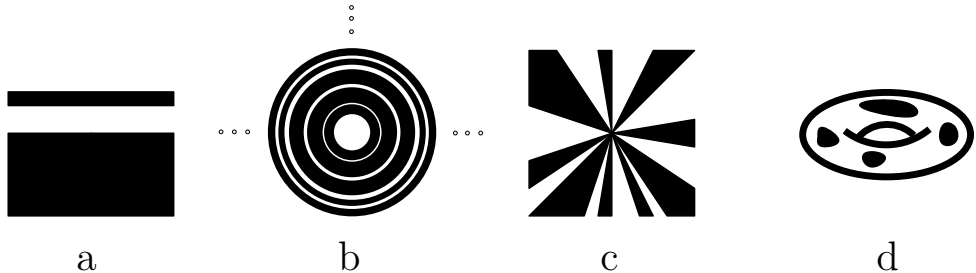}\caption{Four configurations with $\zav\ne1/2$ classified by their symmetry on the $(x_1,x_2)$ plane. (a) Translational symmetry along $x_1$, (b) Rotational symmetry (c) Scaling symmetry and (d) Translational symmetry in both of $x_1$ and $x_2$ directions, in which case one might compactify \xplane\ on a two torus.}\label{Four-Figs}
\end{center}
\end{figure}

\subsection{Asymptotically plane-wave LLM geometries}
As the first case of $\zav\ne1/2$ we consider asymptotically plane-wave geometries for which $\zav=0$. This case and the case discussed in section 3.1 are the only two examples which can be understood as (finite) half BPS deformations of a maximally supersymmetric background, that is the $AdS_5\times S^5$ and the plane-wave backgrounds.
Here again we first identify the good quantum numbers characterizing this class of solutions. For this purpose, similarly  to the previous cases, we study the large $y$ behavior of $z$. The expression to begin with is (\ref{z-strips}) but we should keep in mind that the integral is now over the infinite {\it background} black sea plus the black strips. Choosing the origin of $x_2$ to be on the edge of the sea, the expression for $z$ reads
\be\label{z-pp-large y}
z=\frac{x_2}{2\sqrt{x_2^2+y^2}}-\frac{y^2}{2}\int_{\cal S}dx'_2\frac{1}{((x_2-x'_2)^2+y^2)^{3/2}}\ .
\ee
For finitely extended strips in $x_2$ direction, one might perform the large $y$ expansion of $z$:
\be\begin{split}
z &=\frac{x_2-\Delta}{2y}+\frac{3}{4y^3}\big[-\frac{(x_2-\Delta)^3}{3}-2x_2K_1+K_2\big]
+{\cal O}(\frac{1}{y^5})\cr
&= \frac{x_2-\Delta}{2\sqrt{(x_2-\Delta)^2+y^2}}+\frac{3}{4y^3}(K_2-2x_2K_1)+
{\cal O}(\frac{1}{y^5})\ , 
\end{split}
\ee
where
\begin{subequations}
\begin{align}
\Delta &\equiv \int_{\cal S}dx'_2\ ,\\
K_1&\equiv \int_{\cal S}dx'_2\ x'_2-\frac{\Delta^2}{2}\ ,\\
K_2&\equiv \int_{\cal S}dx'_2\ {x'}_2^2-\frac{\Delta^3}{3}\ .
\end{align}
\end{subequations}
There are a number of points worth mentioning about the above expansion.

$\bullet$ There is no constant term and the expansion starts as $1/y$, reflecting the fact that $\zav=0$.

$\bullet$ As has been made explicit in the second line of (\ref{z-pp-large y}),  the large $y$ expansion takes a simple form using the expansion of  $\frac{x_2-\Delta}{2\sqrt{(x_2-\Delta)^2+y^2}}$ which is the $z$ function for an infinite black sea with the edge at $x_2=\Delta$. This will be useful in finding the physical interpretation of $K_1, K_2$.

$\bullet$ Because of the black sea, there is a preferred origin for $x_2$ and therefore, unlike the strip case, the integral appearing in the expression for $K_1$ does include physical information that cannot be removed by a coordinate transformation.

 We can now read off the quantum numbers of the solution, $\Delta$ and $K_1$, which are in fact the zeroth and first moments of the perturbations around the plane-wave solution respectively.
The width $\Delta$ becomes quantized ({\it cf.} discussions of section 3.2) once we compactify the $x_1$ direction on a circle of radius $R$. The geometry described by the $z$ in this case, then corresponds to (the near horizon geometry) of a stack of $k=\frac{R}{2\pi^2 l_p^4} \Delta$
spherical three brane giants smeared along the $x_1$ direction probing the background plane-wave.

In this case, upon compactification of $x_1$ on a circle, one may perform a light-cone quantization of the string theory on this background. This configuration of strips then corresponds to a specific state of the DLCQ in the sector with light-cone momentum $K_1$, explicitly, $K_1$ which is the first moment of the distribution of strips, may be identified with $p^+$ of the DLCQ theory. This can be seen from the metric and the radii of the three spheres there and the fact that
performing the analysis of stability of spherical  branes probing the background plane-wave, similarly to \cite{Hedge-hog}, one finds that the radius squared of the giant three brane gravitons is proportional to $p^+$. In \cite{TGMT} a matrix theory formulation of DLCQ of type IIB string on the plane-wave has been proposed and in \cite{Torab} it was shown that
the half BPS sector of the tiny graviton matrix theory can be identified with configurations of strips (or Young tableaux of $K_1$ number of boxes). One of the outcomes of the tiny graviton matrix theory is the fact that the \xplane\ is indeed a noncommutative cylinder \cite{progress}.

To get a better feel of what $K_1$ is, let us compute it for the example shown in Fig.\ref{pp+st} (a)
\be
K_1=\int_a^{a+\Delta}dx'\ x'-\frac{\Delta^2}{2}=a\Delta\ .
\label{K1}
\ee
The generalization of the above result to multi strips with $(a_i,\Delta_i)$, where $a_i$ and $\Delta_i$ are respectively the width of the $i^{\rm th}$ successive white and black strips,  is straightforward:
\[
K_1=\sum_i \sigma_i\Delta_i\ ,\ \ \ \sigma_i=\sum_{j=1}^{i} a_j\ .
\]
In the language of the Young tableaux corresponding to the configuration of these strips \cite{LLM, Torab} $K_1$ is nothing but the total number of boxes in the tableau which, as discussed in \cite{Torab}, is equivalent  to the light-cone momentum $p^+$.

\begin{figure}[ht]
\begin{center}
\includegraphics[scale=.9]{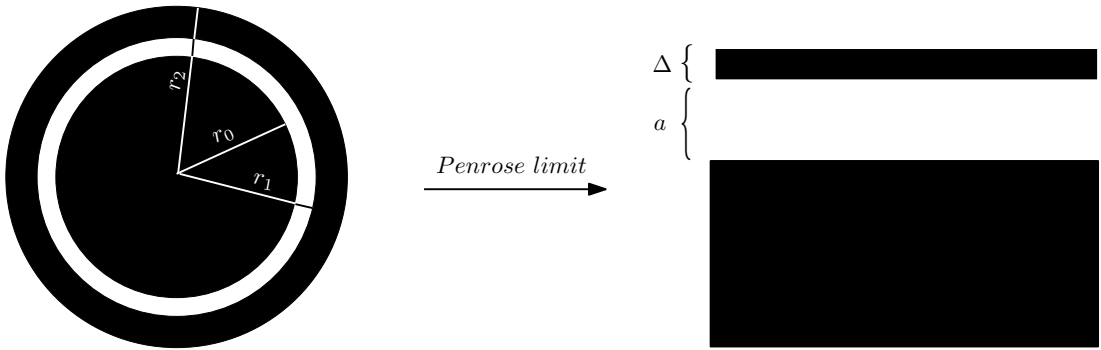}\caption{The rings configurations with $\zav= \half$  and strips configurations of $\zav=0$ are related by Penrose limit. That is, Penrose limit changes the structure of the causal boundary, as noted in \cite{BN}. The expression for $J$ of the rings then directly goes over to $K_1$ of the strips.}
\label{pp+st}
\end{center}
\end{figure}

One may also think about $K_1$ noting that the strips configuration can be obtained as the Penrose limit of multi concentric ring configuration \cite{LLM,{Ebrahimi}}. For the single strip this has been depicted in Fig.\ref{pp+st}. It is interesting to compute $J$ for the latter and compare it with $K_1$
\be
J=\frac{1}{16\pi^2 l_p^8}(r_2^2-r_1^2)(r_1^2-r_0^2)\ .
\ee
While $J$ is proportional to the product of the areas of black and white rings, $K_1$ is proportional to the product of the widths of  black and white strips. If we compactify $x_1$ on a circle of radius $R$, $K_1$ is also proportional to the area of white strip times the area of the black strip. One can also directly apply the Penrose limit to expression for $J$ to obtain $K_1$. To see this it is enough to recall that Penrose limit amounts to \cite{Ebrahimi}
\be\label{Penrose-limit}
r_0\to\infty\ ,\ \ {r_1-r_0}=\frac{a}{r_0}\ ,\  {r_2-r_1}=\frac{\Delta}{r_0}\ ,\ \   a, \Delta={\rm fixed}\ .
\ee

The dual gauge theory and 2d fermion picture for this case has been analyzed in \cite{LLM, LM}
and we skip that here.

\subsection{Configurations with scaling symmetry on \xplane}

In this section we study configurations with scaling symmetry on the \xplane, that is
\be\label{scaling-symmetry}
z(\lambda x_1,\lambda x_2; 0)=z(x_1, x_2; 0)\ .
\ee
These configurations have been discussed in \cite{Shepard}. It is then immediate, using \eqref{V,G,z} and \eqref{z-sol'n},  to check that
\be\label{z-scaling}
z(\lambda x_1,\lambda x_2; \lambda y)=z(x_1, x_2; y)\ ,\ \
V_i( \lambda x_1,\lambda x_2; \lambda y)=\frac{1}{\lambda} V_i(x_1, x_2; y)\ ,
\ee
and hence $ds^2\to \lambda ds^2$.
One should, however, note that this scaling symmetry in $z$ is a classical one and has anomaly. This is due to the fact that \xplane\ is Moyal plane and for fixed $l_p$, \eqref{NCy} breaks the scaling symmetry. The classical configurations which exhibit the scaling symmetry \eqref{scaling-symmetry} are of the form of wedges depicted in Fig.\ref{Wedge-Fig} (a). The origin $x_i=0$ which is the fixed point of the scaling $x_i\to\lambda x_i$ is a ``singular''
point in the sense that on the quantum (Moyal) plane one can never focus on a given point with infinite precision. At quantum level, however, this ``singularity'' is resolved by quantum effects. The two possibilities for this resolution is depicted in Fig. \ref{Wedge-Fig} (b1, b2).
\begin{figure}[h]
\begin{center}
\includegraphics[scale=0.6]{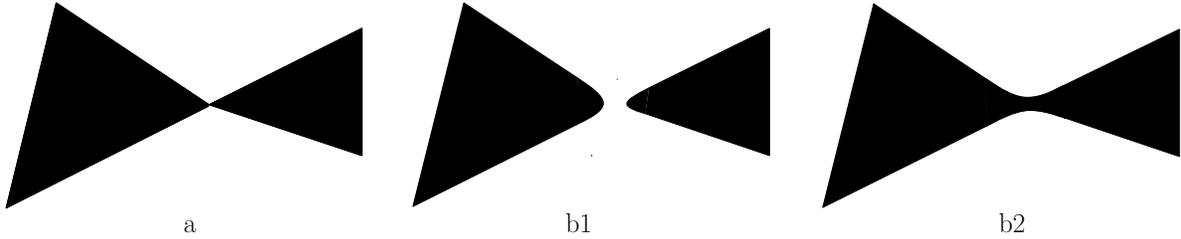}\caption{A generic configuration with scaling symmetry. In  Fig. (a) there is a ``classical'' configuration and in (b1) and (b2) there are two possibilities for ``quantum'' resolution, in which the sharp distribution at $x_i=0$ has been smoothed out.}
\label{Wedge-Fig}
\end{center}
\end{figure}

Let us now focus on the $z$ function for these configurations to read the quantum numbers.
We use polar coordinates $(r,\phi)$ on the plane in terms of which $z$ can be written as
\be
z=\half-\frac{y^2}{\pi}\int_{\cal D}d\phi'\ dr'\ r'\frac{1}{(r^2-2rr'\cos(\phi-\phi')+r'^2+y^2)^2}\ .
\ee
Since we are interested in the large $y$ behavior of the $z$ function we can safely take
$r$ also to be large (compared to $l_p^2$) and hence ignore the ``quantum'' effects and the fact that the scaling symmetry is not exact.\footnote{Of course
noting \eqref{z-scaling} at small and large $y$'s values of $z$ are essentially the same.
What we mean by large $y$ expansion in this case is then considering the $l_p^2\ll r\ll y$
and expanding in powers of $r/y$.} One can then use the scaling symmetry to perform integration over $r'$ to obtain\footnote{It is interesting to note that for the wedge configurations in general one has
$z=\zav+ \epsilon_{ij} x_i V_j$.}
\be
\begin{split}
z&=\half-\frac{y^2}{\pi}\int_{\cal W}d\phi'\biggl[\frac{1}{2(r^2+y^2)}+\half\frac{r^2\cos^2(\phi-\phi')}{(r^2+y^2)(r^2\sin^2(\phi-\phi')+y^2)}+\\
&+\half\frac{r\cos(\phi-\phi')}{(r^2\sin^2(\phi-\phi')+y^2)^{3/2}}\bigl(\frac{\pi}{2}+\tan^{-1}\frac{r\cos(\phi-\phi')}{\sqrt{r^2\sin^2(\phi-\phi')+y^2}}\bigr)\biggr]\ ,
\end{split}
\ee
where the integral ${\cal W}$ is over black wedges. We can now make the large $y$ expansion for $z$
\be\label{z-wedges-expanded}
z=\half-\frac{\Omega}{2\pi}-\frac{r\cos\phi}{4y} {\cal L}_1- \frac{r^2}{4\pi y^2}\big[
2\cos\phi {\cal L}_1+\cos 2\phi {\cal L}_2+\sin 2\phi {\cal L}_2'\big] +
{\cal O}(\frac{r^3}{y^3})\ .
\ee
In the above expression
\begin{subequations}\label{wedges-moments}
\begin{align}
\Omega &=\int_{\cal W}d\phi'\ ,\\ 
{\cal L}_1 &= \int_{\cal W}d\phi'\ \cos\phi'\ ,\\  
{\cal L}_2 &= \int_{\cal W}d\phi'\ \cos 2\phi'\ ,\ \
{\cal L}_2'= \int_{\cal W}d\phi'\ \sin 2\phi'\ ,
\end{align}
\end{subequations}
and we have chosen the origin of the angular coordinate such that $\int_{\cal W}d\phi'\sin\phi'=0$.

The leading term is the zeroth angular moment which is clearly equal to $\zav=\half-\frac{\Omega}{2\pi}$. It is evident that $\zav\ne \half$, unless
$\Omega$ is vanishing. Therefore, the wedges fall into the class of configurations with one dimensional null boundary. As we see  \eqref{z-wedges-expanded} contains all powers, even and odd, of
${r}/{y}$. This may be compared with the case of rings or strips where we only have
even and odd powers of $1/y$, respectively.

Unlike the previous cases one can have wedge configurations where
the zeroth and/or  first leading terms vanish, \ie
$\Omega=\pi$ and/or ${\cal L}_1=0$  and hence the leading order can become $\O(\frac{r^2}{y^2})$.
Examples of this case have been depicted in Fig.\ref{Wedges-Special} .
\begin{figure}[h]
\begin{center}
\includegraphics[scale=0.5]{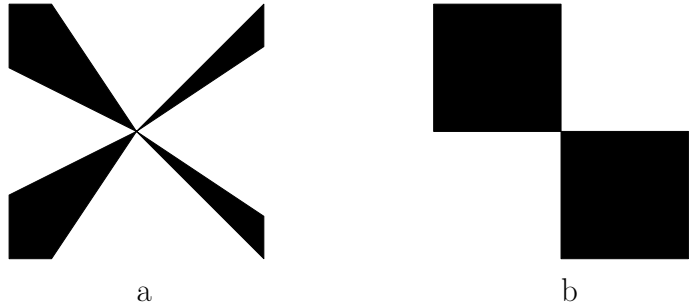}\caption{Two special cases of the wedge configurations.
In Fig. (a) there is a configuration with $\zav\neq 0$ and ${\cal L}_1= 0$. In Fig. (b)
there is a configuration with $\Omega=\pi$ and ${\cal L}_1=0$ for which the large $y$ expansion  starts at $r^2/y^2$ order. This case has been studied in \cite{Shepard}.}\label{Wedges-Special}
\end{center}
\end{figure}
 Another special and interesting class of these configurations
are those which are  invariant under the $Z_2$ symmetry which
exchanges the black and white regions and hence they all have
$\zav=0$. The plane-wave background is a special case of this
kind. One may also recognize the class of configurations   which
keep a discrete subgroup of the $U(1)$ rotation symmetry of the
\xplane . The overlap of the latter two classes are the
configurations which are composed of $2N$ successive black and
white wedges of opening $\pi/N$. This configuration keeps a $Z_N$
subgroup of the rotations. Since the configuration is also
invariant under the black/white exchange $Z_2$ symmetry, the
symmetry of this case enhances to $Z_{2N}$. Despite of the fact
that the wedge configurations receive quantum correction and the
scaling symmetry is anomalous,  this $Z_{2N}$ symmetry can be exact.
This is due to the fact that \eqref{NCy} is invariant under $U(1)$
rotations. An example of this kind has been shown in
Fig.\ref{Wedges-Zn}. One may then consider $Z_N$ or $Z_{2N}$
orbifolds of \xplane . The quantum effects discussed earlier 
in this section will then resolve the orbifold singularity. This
provides a nice and simple example of how stringy/quantum effects
can resolve singularities.
\begin{figure}[h]
\begin{center}
\includegraphics[scale=1]{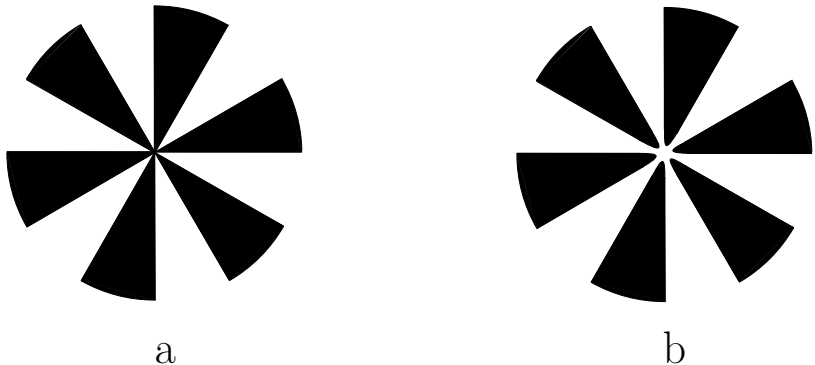}\caption{An example of the wedge configuration with $Z_{2N}$ symmetry and one of its possible quantum mechanically ``resolved'' versions. In this figure $N=3$. Note that
$Z_{2N}=Z_2\times Z_N$, where the $Z_2$ part is the black/white exchange symmetry and the $Z_N$ part is coming from the rotations on \xplane . As we see both of the left and right figures
exhibit the $Z_{2N}$ symmetry.}\label{Wedges-Zn}
\end{center}
\end{figure}

\subsection{ The cases with \xplane\ on a torus}

As the last example of the cases with one dimensional light-like
boundary (that is the cases with $\zav\neq \half$) we consider the
case where the $(x_1,x_2)$ space is a 2-torus with radii
$(R_1,R_2)$. For this we take the distribution on the $(x_1,x_2)$
plane to be periodic in both directions with periodicity $(2\pi
R_1,2\pi R_2)$: \be\label{periodic-z0} z(x_1+2\pi R_1, x_2+2\pi
R_2; 0)=z(x_1,x_2; 0)\ .
\ee

This is a distribution
with infinite extent in both directions as opposed to the finite
extent for asymptotically $AdS_5\times S^5$ and the strips
configurations with infinite extent in one direction, discussed in
section 3. The main point here is to identify the relevant quantum
numbers describing a given configuration. The first point to note
is that the periodicity on the plane results in the periodicity
for $z$ at $y\ne0$ as can easily be checked 
\[
\begin{split}
z(x_i+2\pi
R_i ;y) &=\frac{y^2}{\pi}\int\frac{z(x_i';0)dx'_1dx'_2}{[(x_i+2\pi
R_i-x'_i)^2+y^2]^2}= \frac{y^2}{\pi}\int\frac{z(x_i'+2\pi
R_i ;0)dx'_1dx'_2}{[(x_i-x_i')^2+y^2]^2} = z(x_i ; y)\ ,
\end{split}
\]
where we have used \eqref{periodic-z0}.

Because of the periodicity
it is most natural to make a Fourier expansion of the distribution
on the plane and as we will see the good quantum number(s) for
this case should be sought for among these Fourier modes. Let
us first consider a general  distribution and instead of Taylor expansion in powers of $1/y$, Fourier transform it such that it is applicable to the previous cases. We
will then focus on the periodic distribution.  Consider a
configuration which is given by $z(x_1,x_2,0)$ for which we can
write
\be
z(x_1,x_2,0)=\int z_0(p,q)e^{ipx_1}e^{iqx_2}dp\ dq\ .
\ee
The modes $z_0(p,q)$ can be read as
\be
z_0(p,q)=\frac{1}{(2\pi)^2}\int z(x_1,x_2,0)e^{-ipx_1}e^{-iqx_2}dx_1dx_2\ .
\ee
Now plug this expansion in the expression for $z$
\be
z(x_1,x_2,y)=\int dpdq
z_0(p,q)e^{ipx_1}e^{iqx_2}I(p,q,y)\ ,
\ee
where \cite{Integral-booklet}
\be
I(p,q,y)=\frac{1}{\pi}\int
dudv\frac{e^{ipyu}e^{iqyv}}{(1+u^2+v^2)^2}=y\sqrt{p^2+q^2}K_1(y\sqrt{p^2+q^2})\ ,
\ee
where $K_1(x)$ is the modified Bessel function. For
future use we write the asymptotic behavior of this
function%
\be\label{asymptotic-Bessel}
xK_1(x)\approx\left\{
\begin{array}{cc}
1 \,\,\,\,\,\,\,\,\,\,\,\,\,\,\,\,\,\,\,\,\ \ \ \ \ \ x\ll 1\ ,\cr\;\\
\sqrt{\frac{\pi}{2}x}e^{-x}\ \ \ \,\,\,\ \ x\gg1\ .
\end{array}
\right.
\ee%
Now
consider a finite extent distribution. We know from the analysis
of the previous sections that at large values for $y$ the leading
term for $z$ is the average value on the plane, 1/2, and the
subleading terms form an expansion in powers of $1/y^2$. We can
reproduce these results by the Fourier analysis of this section as
follows. Let us take the following Gaussian distribution
on the plane
\be
z(x_1,x_2,0)=\frac{1}{2}-\exp(-\frac{x_1^2+x_2^2}{l^2})\ .
\ee
This does not exactly produce an allowed LLM boundary condition but we
take it as an approximation to a disk with radius $\sim l$. We now
Fourier expand the second term on the right hand side and plug it
in the expression for $z$. We find that at large values for $y$
\be
z(x_1,x_2,y)\approx\frac{1}{2}-\frac{l^2}{2^{5/2}\pi^{1/2}}\int
dp\ dq
e^{ipx_1}e^{iqx_2}(y\sqrt{t})^{1/2}\exp(-\frac{tl^2}{4})\exp(-y\sqrt{t})\ ,
\ee
where $t=p^2+q^2$. Now if we set $x_1=x_2=0$, we find that
\be
z\approx\frac{1}{2}-\frac{\pi^{1/2}}{2^{3/2}}\frac{l^2}{y^2}\int
du\ u^{3/2}\exp[-(u+\frac{u^2l^2}{4y^2})]\ .
\ee
Note that the dominant contribution to the integral comes from $u<y/l$ and
therefore one can make an expansion in $(\frac{ul}{y})^2$. Thus we
see that for this configuration the large $y$ behavior is as we
expected. The above analysis can of course be repeated for all the previous cases.

We now turn to the case of torus where because of periodicity the Fourier expansion is a discrete one and one has
\be
z(x_1,x_2,0)=\sum_{m,n}z_{mn}e^{\frac{imx_1}{R_1}}e^{\frac{inx_2}{R_2}}\ ,
\ee
where
\be
z_{mn}=\frac{1}{(2\pi)^2R_1R_2}\int dx_1dx_2\ z(x_1,x_2,0)e^{\frac{-imx_1}{R_1}}e^{\frac{-inx_2}{R_2}}\ ,
\ee
and the integration is over the fundamental region. We use this expansion to compute $z$
\be
z(x_1,x_2,y)=\sum_{m,n}z_{mn}e^{\frac{imx_1}{R_1}}e^{\frac{inx_2}{R^2}}I_{mn}(y)\ ,
\ee
where
\be
I_{mn}(y)=y\sqrt{t}K_1(y\sqrt{t})\ ,
\ee
with
\be
t=\frac{m^2}{R_1^2}+\frac{n^2}{R_2^2}\ .
\ee
In the large $y$ limit and at $x_1=x_2=0$ this expression is approximated by
\be\label{expon-fall-off}
z\approx z_{00}+\sum_{m,n\ne0}\ z_{mn} \sqrt{\frac{\pi}{2}}(y\sqrt{t})^{1/2}\exp(-y\sqrt{t})\ .
\ee
Note that we have separated the $z_{00}$ mode because for this mode $t=0$ and the large $y$ approximation for $K_1$ does not work. Instead, for this single mode, we must use the approximate expression of $K_1$ for $y\sqrt{t}\ll1$. The end result is that for large $y$ all but the zero mode are suppressed exponentially. This behavior is different from what we saw in the previous cases and a single mode, $z_{00}$, becomes distinct. This is nothing but the average value of $z$ on the plane or the zeroth moment of the distribution and we identify it as the relevant quantum number for such distributions. For the case where in a basic cell of the torus we have $K$ units of the white region and $N$ units of the black region, \ie
\be\label{volume}
  R_1R_2= l_p^4 (N+K)\ ,
\ee
the average $z$ is 
\be\label{z-average-torus}
\zav=z_{00}=\frac{1}{2}\ \frac{K-N}{K+N}\ ,
\ee
which is not $\half$ unless $N=0$. Obviously the $Z_2$ which exchanges black and white
regions appears as $N\leftrightarrow K$ symmetry. In the corresponding quantum Hall
terminology, despite the fact that $\half-\zav=\frac{N}{N+K}$ gives the density of the particles which is generically not equal to zero or one,
microscopically we still   have an integer (as opposed to fractional) quantum Hall system \cite{LLLvsLLM}.

\begin{figure}[t]
\begin{center}
\includegraphics[scale=.7]{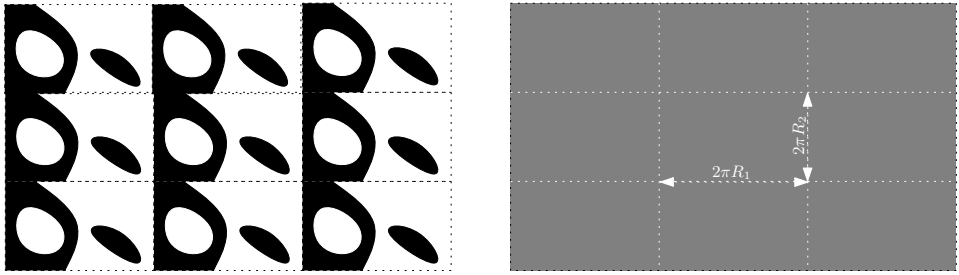}\caption{A generic configuration on the torus lattice. The left figure shows the black and white distribution and the right figure is the same distribution averaged over. This has been depicted using a  gray color. A similar color coding was also used in \cite{Vijay, Alishah, LM}.}
\label{period}
\end{center}
\end{figure}

Finding the dual field theory for these configurations requires some care. As in section
3.1 we use the spherical threebrane  probes. In this case, however, to simplify the picture
we consider a limit where the background is essentially looking like a ``gray'' background, as depicted in right figure of Fig.\ref{period}. This is basically when our probes are viewing the
\xplane\ at large $y$. The full picture is more complicated  and to analyze that we need to really consider the little string theory \cite{LM}. 
In this case one has the option of using 
either the giant or the dual giant  probes. In the probe approximation, the system is described by the spherical branes represented by the distribution in a single cell to be the objects which probe the background created by the rest of the distribution on the \xplane. So our probes are either $N$ branes wrapping $\Omega_3$ (dual giants), represented by the black area, or $K$ branes wrapping $\tilde{\Omega}_3$ (giants) represented by the white area,  as  depicted in Fig.\ref{gray-probe}. The dual giant gravitons 
(giant gravitons) description is a good one in the $N\ll K$ ($K\ll N$) regimes. 

To obtain the 
dual theory we need to refine analysis of section 3.1. In order that we need to have the relevant four-forms in the background. In the large $y$ limit where $z=\langle z\rangle$  (recall the exponential fall-off in $z-\langle z\rangle$, {\it cf.} \eqref{expon-fall-off}), the part of the background RR four-form which is relevant to the spherical brane probes is \cite{LLM}
\be\label{C4-background}
\begin{split}
C_{4}=A\wedge d\Omega_3&+{\cal A}\wedge d\tilde{\Omega}_3\ ,\\
A=\frac{1}{4}( \langle z\rangle+\half)\ \epsilon_{ij}x^idx^j &,\ \ \  
{\cal A}=\frac{1}{4}( \langle z\rangle-\half)\ \epsilon_{ij}x^idx^j\ ,
\end{split}
\ee
where in the torus case $\epsilon_{ij}$, which is standing as the volume form (or Kahler form) on the torus, is proportional to $\sqrt{\det g}={\frac{R_1R_2}{l_p^4}}={N+K}$ \eqref{volume}. 
\begin{figure}[t]
\begin{center}
\includegraphics[scale=.8]{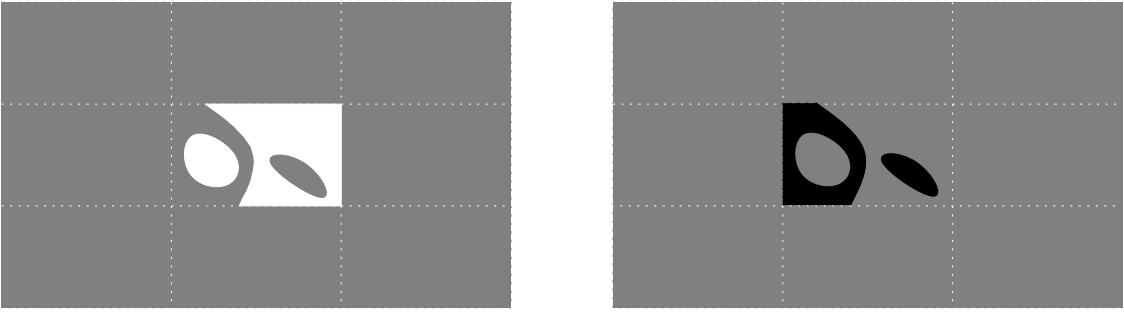}\caption{The dual gauge theory description of the toroidal configuration given by the average $\zav\neq \half$, which has been depicted as the gray background, can be either through the theory of giant threebranes (the right figure) or that of the dual giant threebranes  (the left figure).}\label{gray-probe}
\end{center}
\end{figure}
For the $N$ giant probes (or $K$ dual giant probes) the theory is  then a quantum Hall system in the background magnetic field $B$  (or ${\cal B}$) where
\be\label{B-fields}
\begin{split}
B=\frac{1}{2}( \langle z\rangle+\half)\ (N+K) &=\frac{1}{4}K\ , \\
{\cal B}=\frac{1}{2}( \langle z\rangle+\half)\ (N+K) &=-\frac{1}{4}N\ . 
\end{split}
\ee
Repeating the analysis of section 3.1 the theory of this probes is $U(N)$ Chern-Simons Matrix theory at level $K$ (or $U(K)$ Chern-Simons Matrix theory at level $N$) on the two torus, that is:
\begin{subequations}\label{U(N)K}
\begin{align}
S&=\frac{K}{4\pi}\int dt \  Tr_{N}(\epsilon_{ij} X^iD_0X^j)\ ,\\
U_j X_i U_j^{-1}&=X_i+2\pi\delta_{ij} R_i\ ,\ \ \ i,j=1,2\ .
\end{align}
\end{subequations}
This is the result discussed in \cite{LM}.  The fact that the coordinates on the torus
do not commute, as in \eqref{NCy},  results from the fermionic nature of the droplets or incompressibility of the corresponding quantum Hall liquid \cite{LLLvsLLM}. It is worth noting that although here we are considering the toroidal case, the above arguments also 
hold for a generic black and white configuration, e.g. consisting of (infinitely) many droplets, with $\zav\neq \half$.

The two theories, $U(N)_K$ and $U(K)_N$ are then related by the $Z_2$  black and white exchange symmetry and hence one would expect them to be equivalent, the level rank duality. This has indeed been shown and discussed in \cite{level-rank}. This suggests that, although 
we obtained $U(N)_K$ in the $N\ll K$ limit, it should be a good description for generic $N, \ K$
\cite{LM}.

The geometry described by the configuration depicted in Fig.\ref{period} (the gray \xplane )
is the near horizon geometry of $K$ giants and $N$ dual giants smeared uniformly in the $x_1$ and $x_2$ directions.  As discussed in \cite{LM} one may perform two T-dualities and an S-duality on the above brane configuration. This geometry then goes over to (the near horizon geometry of) intersection of NS5branes of type IIB $N$ of which have worldvolume along $(t,x_1,x_2, \Omega_3)$ and $K$ of them along $(t,x_1,x_2,\tilde \Omega_3)$. Again in the $N\ll K$ limit one can take the $K$ fivebranes as background and the stack of $N$ as probes which in the 1/2 BPS sector this leads to $(2+1)$ dimensional $U(N)_K$ Chern-Simons gauge theory on the the dual torus.

One may follow the above dualities directly at the level of the Matrix Chern-Simons theory. 
As it is well-known from the BFSS Matrix theory literature \cite{BFSS}
\be\label{BFSS-T-dual}
{\rm Matrix\ theory \ (0+1)\ gauge\ theory}/T^2\equiv
{\rm (2+1)\ gauge \ theory}/\tilde T^2\ ,
\ee
where $\tilde T^2$ is the torus dual to $T^2$. Hence, recalling that Chern-Simons is a topological theory, starting from a Matrix $U(N)_K$ Chern-Simons theory on the $T^2$
one obtains a $(2+1)$ dimensional $U(N)_K$ Chern-Simons theory on the dual torus $\tilde T^2$.

Of course the above statement can be written in a more general way. Recall that the T-duality group on the torus is $SL(2,Z)_\tau\times SL(2,Z)_\rho$ where the $SL(2,Z)_\rho$ is acting on the Kahler structure $\rho$ and $SL(2,Z)_\tau$ on the complex structure $\tau$. The $SL(2,Z)_\tau$ is obviously the symmetry of both sides on \eqref{BFSS-T-dual}. The $SL(2,Z)_\rho$ is, however, non-trivial and in  \eqref{BFSS-T-dual}  only a $Z_2\in SL(2,Z)_\rho$, which maps the pure imaginary  $\rho$ to a pure imaginary $\rho$,  has become manifest. Generically $SL(2,Z)_\rho$ relate the Matrix theory to a $(2+1)$ theory on a noncommutative torus $T^2_{\Theta}$ \cite{CDS}
\be\label{NC-torus-duality}
 {\rm Matrix\ theory \ (0+1)\ gauge\ theory}/T^2\equiv
{\rm (2+1)\ gauge \ theory}/\tilde T^2_{\Theta}\ ,
\ee
where $SL(2,Z)_\rho$ on $\Theta$ act as $\Theta\to\frac{a\Theta+b}{c\Theta +d}$, $ad-bc=1,\ a,b,c,d \in {\mathbb Z}$. In the language of probes $U(N)_K$ Chern-Simons on $T^2_{\Theta}$ 
 is the theory on (non-marginal) bound state of $N$ (NS5,D3) branes probing another
 (NS5,D3) bound state. We would like to stress that the black/white $Z_2$ exchange symmetry is not a part of the T-duality group $SL(2,Z)_\rho$ and is a symmetry which becomes manifest only in the 1/2 BPS sector and in the corresponding LLM geometries.
 
Although the above $(2+1)$ dimensional $U(N)_K$ theory provides a generically good description,
there are some specific black and white configurations on the torus  where a $(1+1)$ dimensional
description may become a perturbative description. This is the case where for a given $N,\ K$
the black region is like a narrow strip, as depicted in Fig. \ref{hierarchy}. In this case
one can approximate the system with a system of strips which we discussed in  section 3.2.
In the gravity picture, this is basically performing one T-duality. In the dual
gauge (0+1) theory (which is the 1/2 BPS sector of a $3+1$ dimensional gauge theory) 
this can also be understood through the ``deconstruction'' phenomena \cite{deconstruct}.
\begin{figure}[ht]
\begin{center}
\includegraphics[scale=.7]{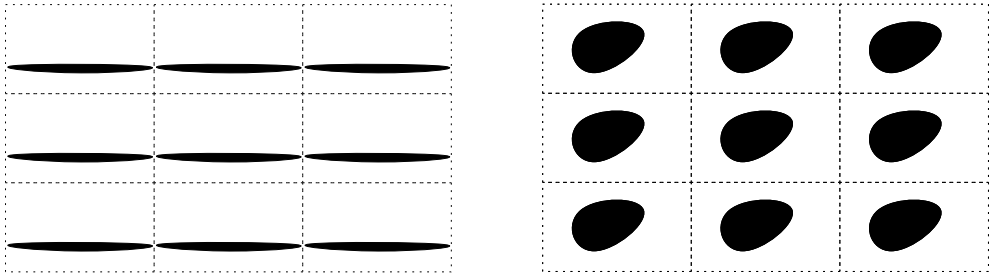}\caption{Examples of distributions on a torus: a distribution with a (1+1) dimensional perturbative field theory description (the left figure) and a generic distribution (the right figure).}
\label{hierarchy}
\end{center}
\end{figure}

\section{Discussion and Outlook}%

In this paper we have studied the ten dimensional LLM half BPS ``bubbling geometries'' \cite{LLM} and tried to classify them. As the ten dimensional LLM geometries are completely specified by the  black/white distribution on the \xplane, $z$, in our classification we focused on the $z$-function. 
As the first criterion we focused on the causal structure of the LLM geometries. Since the LLM geometries are horizon-free non-singular geometries, we concentrated on the structure of the causal boundary and showed that the LLM geometries fall into two classes, those with one dimensional light-like boundary and those with four dimensional $R\times S^3$ boundary.
The latter have $\langle z\rangle=\half$ while the former have $\langle z\rangle \neq \half$. 

In \cite{LLM} a class of half BPS eleven dimensional geometries with $SU(4|2)$ super-isometries have also been discussed. These are geometries governed by the Toda equation and the singularity-free condition leads to two boundary conditions on the \xplane . As we do not know how to obtain a generic solution to the Toda equation, the analysis of the eleven dimensional 
LLM geometries has not been done at the same extent as the ten dimensional case. There are,
however, evidence supporting the idea that the eleven dimensional LLM solutions can also be
described through a distribution of ``black and white'' regions on the \xplane\ \cite{LM}. 
In this case, however, we do not have the black and white exchange symmetry. 
A generalization of our discussions in section 2 indicates that from the causal boundary viewpoint the eleven dimensional LLM geometries fall into \textit{three} classes:
those which are black on the average have $R\times S^2$ boundary, those which are white on the average have $R\times S^5$ boundary and those which are  ``gray'' on the average have one dimensional null boundary.
Establishing the above statement and generalization of some of our results to the eleven dimensional case is among the interesting questions we postpone to future works \cite{progress11}.

It worth noting that in \cite{LM} another classification of the LLM geometries was considered 
which has some  overlaps with ours. In \cite{LM} the \textit{topology} of the \xplane\ was used as the classification criterion, according which the \xplane, which is necessarily a flat two dimensional space in the ten dimensional LLM setup, may be an $R^2$ plane, a cylinder or a torus. Our classification is, however, based on $\langle z\rangle$ and is refined by the zeroth, first and second moments of the $z$-distribution. As we discussed in some detail that is at most the second moment which appears among the {\it global} charges governing the geometries. At the gravity level these are the ADM type charges which correspond to the rank of the gauge group and the global $R$-charge in the dual gauge theory. In sections 3, 4 we discussed in detail which of these moments are the relevant ones for a given distribution.  
It is of course interesting to generalize our  classification criteria or that of \cite{LM}  to the
eleven dimensional LLM solutions. In the eleven dimensional case, unlike the ten dimensional case, the \xplane\ is not necessarily a two dimensional flat space. This, potentially, provides a greater variety of possibilities. Due to the conformal symmetry of the Toda equation \cite{LLM}, one may use the Euler character of the \xplane\ as the base for classification.
However, there seem to be some difficulties with the smoothness of the compact \xplane\ cases with non-vanishing Euler character \cite{unpublished} and hence the smoothness condition forces us to three flat cases of $R^2$ plane, cylinder and torus. A detailed and thorough analysis of this obviously interesting direction is awaiting further studies.

In section 4.2 we discussed $z$-distributions with scaling symmetry and discussed that at ``classical'' level these distributions have a ``singular'' point, the fixed point of the scaling symmetry, and that this singular point is removed by the ``quantum'' corrections and the
fact that the \xplane\ is a noncommutative Moyal plane. In the same class of the LLM geometries 
we discussed cases with $Z_{2N}$ isometries. Performing the orbifolds of this class of geometries we find non-supersymmetric type IIB backgrounds which generically have closed
string tachyons. In these cases, it is usually believed that the orbifold singularity is resolved
when the tachyon is condensed. Here, however, as we discussed the resolution of orbifold  singularity can be understood through the quantum nature of the \xplane\ and that one cannot
probe the \xplane\ with precision higher than $l_p^4$ using the spherical threebrane probes.
It is an interesting open question to address the tachyon dynamics in the orbifolds of this class of LLM geometries.

%
\end{document}